\documentclass[aps,pre,twocolumn,showpacs,superscriptaddress]{revtex4}
\usepackage{graphicx}
\usepackage{bm}

\begin{document}

\title{\boldmath Hot Spots and Transition from $d$--Wave to Another
  Pairing Symmetry in the Electron-Doped Cuprate Superconductors}

\author{V.~A.~Khodel} 
\affiliation{Russian Research Centre Kurchatov
  Institute, Moscow 123182, Russia}
\affiliation{Condensed Matter Theory Center and Center for
  Superconductivity Research, Department of Physics, University of
  Maryland, College Park, Maryland 20742-4111, USA}

\author{Victor M.~Yakovenko} 
\affiliation{Condensed Matter Theory Center and Center for
  Superconductivity Research, Department of Physics, University of
  Maryland, College Park, Maryland 20742-4111, USA}

\author{M.~V.~Zverev}
\affiliation{Russian Research Centre Kurchatov
  Institute, Moscow, 123182, Russia}

\author{Haeyong Kang}
\affiliation{Department of Physics, Ewha Womans University, Seoul
  120-750, South Korea}
\affiliation{Condensed Matter Theory Center and Center for
  Superconductivity Research, Department of Physics, University of
  Maryland, College Park, Maryland 20742-4111, USA}

\date{\bf cond-mat/0307454, v.1: 17 July 2003, v.2: 4 December 2003,
v.3: 13 March 2004}


\begin{abstract}
  We present a simple theoretical explanation for a transition from
  $d$-wave to another superconducting pairing observed in the
  electron-doped cuprates.  The $d_{x^2-y^2}$ pairing potential
  $\Delta$, which has the maximal magnitude and opposite signs at the
  hot spots on the Fermi surface, becomes suppressed with the increase
  of electron doping, because the hot spots approach the Brillouin
  zone diagonals, where $\Delta$ vanishes.  Then, $d_{x^2-y^2}$
  pairing is replaced by either singlet $s$-wave or triplet $p$-wave
  pairing.  We argue in favor of the latter and propose experiments to
  uncover it.
\end{abstract} 
 
\maketitle

\section{Introduction}

The superconducting pairing symmetry in the electron-doped cuprates
\cite{Uchida}, such as $\rm Nd_{2-x}Ce_xCuO_4$ and $\rm
Pr_{2-x}Ce_xCuO_4$, has been debated for a long time.  Originally, it
was thought to be of the $s$-wave type \cite{s-wave}.  Later,
observation of the half-quantum magnetic flux in tricrystals
\cite{Kirtley}, improved microwave measurements of temperature
dependence of the London penetration depth \cite{Kokales}, the
angular-resolved photoemission spectroscopy (ARPES) \cite{Armitage-SC}
and Raman scattering \cite{Blumberg} studies, and observation of the
$\sqrt{H}$ dependence of specific heat on magnetic field $H$
\cite{Balci} pointed to the $d$-wave symmetry.  Recently, evidence was
found for a transition from $d$- to $s$-wave pairing symmetry with the
increase of electron doping \cite{Biswas,Skinta}.  Biswas {\it et
al.}\ \cite{Biswas} concluded that $\rm Pr_{2-x}Ce_xCuO_4$ has
$d$-wave pairing at $x\sim0.15$ and $s$-wave pairing at $x\sim0.17$.
In this paper, we propose a simple scenario for the transition from
the $d$-wave to another pairing symmetry and argue that the latter can
actually be triplet $p$-wave.

\begin{figure}[t] 
\includegraphics[width=0.8\linewidth]{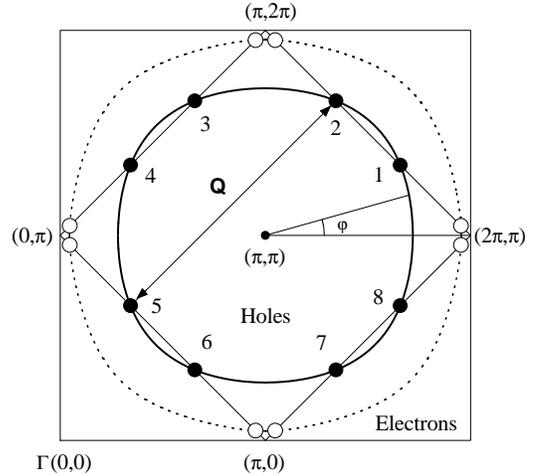}
\caption{Fermi surfaces of Eq.\ (\ref{sp}) for hole doping 
  (dashed line, $\mu=-1.76$, $x=0.48$) and electron doping (solid
  line, $\mu = -0.4$, $x=-0.15$).  The hot spots are shown by open and
  solid circles.  The radius of the circles $\sigma=0.1$ represents
  the width of the interaction (\ref{V_0}) in the momentum space.}
\label{fig:FS}
\end{figure}

First we present a qualitative picture in terms of the Fermi surface
geometry shown in Fig.\ \ref{fig:FS}.  According to the theoretical
model \cite{Scalapino,Pines,Ueda}, the antiferromagnetic spin
fluctuations (ASF) peaked at the wave vector ${\bm Q}=(\pi,\pi)$ are
responsible for $d$-wave superconductivity in the hole-doped cuprates.
Commensurate ASF at the wave vector ${\bm Q}$ are also observed in the
electron-doped cuprates \cite{Yamada}.  The interaction via ASF has
the highest strength at the so-called hot spots, the points on the
Fermi surface connected to each other by the vector ${\bm Q}$.  These
points are labeled in Fig.\ \ref{fig:FS} by the consecutive numbers
from 1 to 8.  Since the interaction via ASF is repulsive in the
singlet channel, the superconducting pairing potential $\Delta({\bm
p})$ has opposite signs at the two hot spots connected by the vector
${\bm Q}$:
\begin{equation}
  \Delta({\bm p}+{\bm Q})=-\Delta({\bm p}).
\label{-Delta}  
\end{equation}
Thus, the eight hot spots can be divided into four groups (1,6),
(2,5), (3,8), and (4,7), with the signs of $\Delta({\bm p})$ being
opposite within each group.  However, the relative signs of
$\Delta({\bm p})$ between the different groups have to be determined
from additional considerations.

In Fig.\ \ref{fig:FS}, the dashed and solid lines show the Fermi
surfaces corresponding to the hole- and electron-doped cuprates.
Notice that the $\Gamma$ point $(0,0)$ is located at the corner of
Fig.\ \ref{fig:FS}, so that the area inside the Fermi surface is
occupied by holes and outside by electrons.  The dashed Fermi surface,
corresponding to the hole-doped case, encloses a larger area, and the
pairs of hot spots shown by the open circles in Fig.\ \ref{fig:FS} are
located close to the van Hove points $(0,\pi)$, $(\pi,0)$,
$(2\pi,\pi)$, and $(\pi,2\pi)$.  It is natural to assume that
$\Delta({\bm p})$ has the same sign within each pair of the
neighboring hot spots.  This assumption, in combination with Eq.\
(\ref{-Delta}), immediately results in the familiar $d_{x^2-y^2}$
symmetry of the pairing potential.

However, the situation does change in the electron-doped case.  With
the increase of electron doping, the Fermi surface shrinks, and the
hot spots move away from the van Hove points toward the Brillouin zone
diagonals.  The following pairs of the hot spots approach each other:
(1,2), (3,4), (5,6), and (7,8).  The $d_{x^2-y^2}$ pairing potential
has opposite signs within each pair and vanishes at the zone
diagonals.  Thus, in the electron-overdoped cuprates, when the hot
spots get close enough, the $d_{x^2-y^2}$ pairing becomes suppressed.
Then, a superconducting pairing of another symmetry may emerge, with
the pairing potential of the same sign on both sides of the zone
diagonals.  This is the mechanism that we propose for the transition
observed in Refs.\ \cite{Biswas,Skinta}.

\section{Suppression of $d$-wave pairing}

To illustrate how the $d_{{x^2-y^2}}$ pairing evolves with doping, we
perform calculations employing the typical electron dispersion law
\begin{equation}
  \xi({\bm p})=-\mu - 2t_0(\cos p_x +\cos p_y)+4t_1\cos p_x\cos p_y
\label{sp}
\end{equation}
with $t_1/t_0=0.45$.  The chemical potential $\mu$ controls the hole
concentration $n$, which is determined by the area $S$ inside the
Fermi surface in Fig.\ \ref{fig:FS}: $n=2S/(2\pi)^2$.  The doping
$x=n-1$ is defined as the deviation of $n$ from half filling, so that
$x>0$ and $x<0$ correspond to hole and electron doping
\cite{x-doping}.  The relation $S\propto1+x$ is in agreement with
ARPES, except for the region of small doping around $x=0$, where the
antiferromagnetic Mott insulating state intervenes.  For $\rm
Nd_{2-x}Ce_xCuO_4$, this was established in Ref.\ \cite{Armitage-FS},
and movement of hot spots toward the zone diagonals with the increase
of electron doping was directly observed in Refs.\
\cite{Armitage-FS,Armitage-PG,Armitage-SC}.  Notice that, for the
dispersion law (\ref{sp}), the hot spots exist only within a finite
range of chemical potential $-4t_1\leq\mu\leq0$, which corresponds to
the range of doping $-0.25=x_-<x<x_+=0.53$.  The respective pairs of
the hot spots merge and disappear at the van Hove points when $x\to
x_+$ and at the zone diagonals when $x\to x_-$.  Thus, in this model,
the $d_{x^2-y^2}$ superconductivity can exist only within a finite
range of electron and hole doping, in qualitative agreement with the
experimental phase diagram of cuprates.  Doping dependence of the
Fermi surface in the electron-doped cuprates obtained from the ARPES
measurements \cite{Armitage-FS,Armitage-PG,Armitage-SC} was
quantitatively interpreted within a simple band-structure model in
Ref.\ \cite{Kusko}.  The results are in qualitative agreement with the
Hall coefficient measurements \cite{Dagan}.

To verify the qualitative picture given in the Introduction, we solve
the BCS equation for the pairing potential
\begin{equation}
  \Delta_{\alpha\beta}({\bm p})= -
  \int V_{\alpha\beta}^{\gamma\delta}({\bm p}-{\bm p}')
  {\tanh{E({\bm p}')\over 2T}\over 2E({\bm p}')}
  \Delta_{\gamma\delta}({\bm p}') {d^2{\bm p}'\over (2\pi)^2}.
\label{bcst}
\end{equation}
Here $E({\bm p})=\sqrt{\xi^2({\bm p})+\Delta^2({\bm p})}$, $T$ is
temperature, and $V_{\alpha\beta}^{\gamma\delta}(q)=
V_c(q)\delta_\alpha^\gamma\delta_\beta^\delta +
V_s(q)\,{\bm\sigma}_\alpha^\gamma\cdot{\bm \sigma}_\beta^\delta$ is
the effective interaction between electron charges and spins, where
${\bm\sigma}$ are the Pauli matrices, and $\alpha,\beta,\gamma,\delta$
are the spin indices.  For singlet and triplet pairings, the functions
$V_0(q)=V_c(q)-3V_s(q)$ and $V_1(q)=V_c(q)+V_s(q)$ enter Eq.\
(\ref{bcst}), respectively.  To simplify our calculations, we ignore
the frequency dependence of $V$ and use the conventional ASF
interaction of the form \cite{Pines}
\begin{equation}
  V_0(q)={g\ \over ({\bm q}-{\bm Q})^2+\sigma^2}
\label{V_0}
\end{equation}
with the coupling constant $g=2t_0$ and the width $\sigma=0.1$
\cite{disclaimer}.

\begin{figure}[t] 
\includegraphics[width=\linewidth]{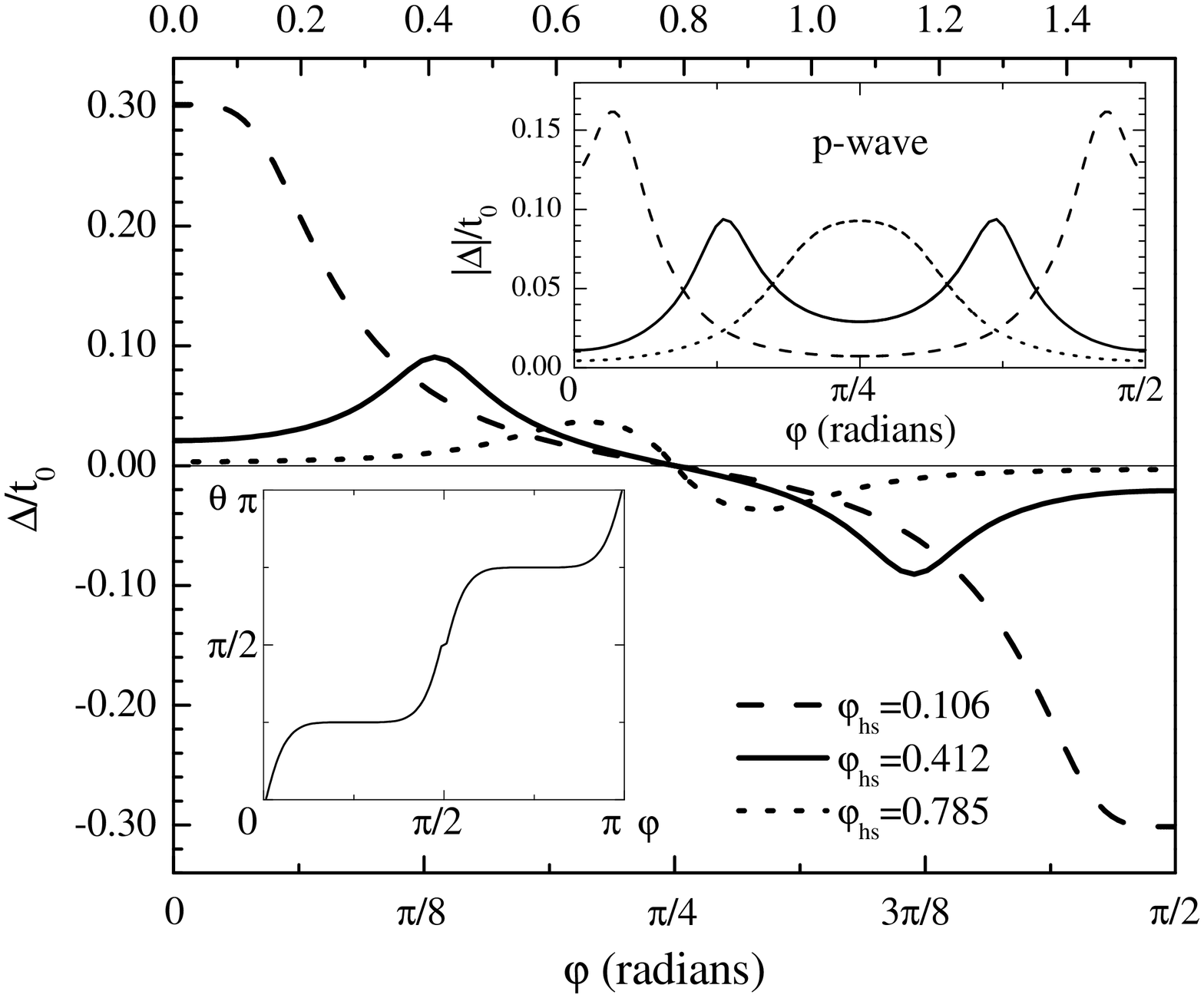}
\caption{
  The pairing potential $\Delta$ at $T=0$ vs.\ the angle $\varphi$ on
  the Fermi surface, shown by the dashed line for $x=0.37$
  ($\mu=-1.6t_0$), the solid line for $x=-0.1$ ($\mu=-0.6t_0$), and
  the dotted line for $x=-0.25$ ($\mu=0$).  The main panel represents
  the $d_{x^2-y^2}$ state, and the upper inset the chiral $p$-wave
  state.  The angle $\varphi_{\rm hs}$ indicates position of the hot
  spot 1.  The lower inset shows the phase $\theta$ of
  $\Delta(\varphi)=|\Delta|e^{i\theta}$ for the chiral $p$-wave
  pairing.}
\label{fig:Delta}
\end{figure}

The $d_{x^2-y^2}$ pairing potential $\Delta$, calculated at $T=0$ for
three different dopings, is shown in the main panel of Fig.\
\ref{fig:Delta} vs.\ the angle $\varphi$ on the Fermi surface (see
Fig.\ \ref{fig:FS}).  The dashed line refers to the strong hole doping
$x=0.37$ close to $x_+$, the dotted line to the strong electron doping
$x=x_-=-0.25$, and the solid line to the intermediate electron doping
$x=-0.1$.  The angle $\varphi_{\rm hs}$ indicates the position of the
hot spot 1 for these dopings.  We see that the maxima of
$|\Delta(\varphi)|$ are achieved at the hot spots, i.e.\ at
$\varphi\simeq\varphi_{\rm hs}$, as discussed in Ref.\ \cite{KY}.  The
solid curve in Fig.\ \ref{fig:Delta} qualitatively agrees with the
nonmonotonic function $\Delta(\varphi)$ inferred from the Raman
scattering in $\rm Nd_{1.85}Ce_{0.15}CuO_4$ \cite{Blumberg}.  We also
observe that $|\Delta|$ drops precipitously when the hot spots
approach the zone diagonals.  This happens because the integral in
Eq.\ (\ref{bcst}) is suppressed when positive and negative peaks of
$\Delta(\varphi)$ are close to each other.

\section{Alternative superconducting pairings}

Once the $d_{x^2-y^2}$ pairing is suppressed in the case of strong
electron doping, pairing of a different symmetry may emerge in the
system.  Evidently, this pairing should provide the same sign of
$\Delta$ within each pair (1,2), (3,4), (5,6), and (7,8) of the
approaching hot spots.  There are three possibilities depending on the
relative signs of $\Delta$ between the different pairs of the hot
spots.  The same sign for all the hot spots corresponds to $s$-wave,
the opposite sign between (1,2) and (3,4) to $d_{xy}$-wave, and the
opposite sign between (1,2) and (5,6) to triplet $p$-wave pairing.  We
need to find out which of these states wins.

Measurements of the temperature dependence of the penetration depth
$\lambda(T)$ show a transition from a gap with nodes to a nodeless gap
with the increase of electron doping in Pr$_{2-x}$Ce$_x$CuO$_4$ and
La$_{2-x}$Ce$_x$CuO$_{4-y}$ \cite{Skinta}.  The point contact
spectroscopy of Pr$_{2-x}$Ce$_x$CuO$_4$ \cite{Biswas} shows a
transition from a strong zero-bias conductance peak, originating from
the midgap Andreev surface states in the $d$-wave case, to double
peaks typical for $s$-wave.  These experiments eliminate $d_{xy}$-wave
pairing, because it has gap nodes and the midgap Andreev states.
$d_{xy}$-wave pairing was proposed in Ref.\ \cite{Guinea} as a
possible successor to $d_{x^2-y^2}$ in the electron-overdoped phase.
In the theoretical model of Ref.\ \cite{Guinea}, nonlocal corrections
to the Hubbard interaction $U$ due to spin fluctuations were taken
into account only in the lowest order in $U$, whereas in our model
(\ref{V_0}) the peak at ${\bm Q}$ is obtained by summing an infinite
number of RPA-like diagrams.  The interaction (\ref{V_0}) peaked at
${\bm Q}=(\pi,\pi)$ is not favorable for $d_{xy}$-wave pairing.

The simplest alternative pairing symmetry consistent with the
experiments \cite{Biswas,Skinta} is $s$-wave, which can be produced by
phonons.  This scenario was proposed by Abrikosov \cite{Abrikosov},
who argued that, with the increase of doping, $d$-wave
superconductivity is destroyed by disorder, whereas $s$-wave
superconductivity survives.  The $s$-wave energy gap $|\Delta|$ has no
nodes and is roughly uniform along the Fermi surface.  However, the
$s$-wave scenario encounters some problems.  When $|\Delta({\bm p})|$
varies along the Fermi surface, measurements of $\lambda(T)$ yield the
minimal value of the gap $\Delta_{\rm min}$ at $T=0$.  The experiment
\cite{Skinta-0.85} found $\Delta_{\rm min}/T_c\simeq0.85$, whereas,
for the phonon-induced $s$-wave superconductivity, this ratio should
be close to the BCS value 1.76.  Furthermore, for the phonon
mechanism, $T_c$ is not expected to depend on doping significantly
\cite{Abrikosov}, whereas the experimental $T_c$ declines steeply at
$|x|\agt0.15$ and vanishes for $|x|\agt0.2$ outside of the dome-shaped
phase diagram of the electron-doped cuprates \cite{Uchida,Peng}.
Incidentally, the value of doping where superconductivity disappears
is close to $x_-$, which indicates that the hot spots may be equally
important for the alternative superconducting pairing.

Thus, it is worth considering the last alternative pairing, namely the
triplet $p$-wave.  It has the order parameter
$\Delta\epsilon_{\alpha\gamma}{\bm \sigma}_\beta^\gamma\cdot{\bm n}$,
where $\epsilon_{\alpha\gamma}$ is the antisymmetric spin tensor, and
${\bm n}$ is the unit vector of spin-polarization \cite{Wolfle}.  The
symmetry of triplet pairing in a tetragonal crystal was classified in
Ref.\ \cite{symmetry}.  In the $\rm E_u$ representation, ${\bm n}$
points along the ${\bm c}$ axis, and the phase of $\Delta({\bm p})$
changes by $2\pi$ around the Fermi surface.  This order parameter is
chiral and breaks the time-reversal symmetry.  The simplest example is
$\Delta({\bm p})\propto(\sin p_x\pm i\sin p_y)$, which was originally
proposed for Sr$_2$RuO$_4$ \cite{Miyake99}.  In the $\rm A_{1u}$, $\rm
A_{2u}$, $\rm B_{1u}$, and $\rm B_{2u}$ representations, the vector
${\bm n}$ lies in the $({\bm a},{\bm b})$ plane and rotates around the
Fermi surface by the angle $2\pi$.  These order parameters are not
chiral and do not break the time-reversal symmetry.  Both types of the
pairing potential have two components $(\Delta_1,\Delta_2)$, the real
and imaginary parts of $\Delta$ in the chiral case and
$(n_x\Delta,n_y\Delta)$ in the nonchiral case, which satisfy the
symmetry relation $|\Delta_2(p_x,p_y)|=|\Delta_1(p_y,p_x)|$.  Then,
the gap $|\Delta|^2=\Delta_1^2+\Delta_2^2$ does not have nodes, but is
modulated along the Fermi surface.  This easily explains the reduced
value of $\Delta_{\rm min}/T_c$ observed in Ref.\ \cite{Skinta-0.85}.
The tunneling spectrum, shown in Fig.\ 3 of Ref.\ \cite{Sengupta} for
$\Delta\propto(\sin p_x\pm i\sin p_y)$, has double peaks, as in the
experiment \cite{Biswas}.  Thus, the experiments
\cite{Biswas,Skinta,Skinta-0.85} are compatible with both $s$- and
$p$-wave pairings and are not sufficient to distinguish between them.

Measurements of the Knight shift can distinguish between singlet and
triplet pairing.  The Knight shift in the electron-doped $\rm
Pr_{0.91}LaCe_{0.09}CuO_{4-y}$ was found to decrease below $T_c$
consistently with the singlet $d$-wave pairing \cite{Zheng}.  However,
the Knight shift in $\rm Pr_{1.85}Ce_{0.15}CuO_{4-y}$ was found not to
change below $T_c$ \cite{Zamborszky}.  This is an indication of
triplet pairing, like in Sr$_2$RuO$_4$ \cite{Ishida98} and in the
organic superconductors $\rm(TMTSF)_2X$ \cite{Chaikin}.  To obtain a
complete picture, it is desirable to measure the Knight shift in the
superconducting state systematically as a function of electron doping
across the transition from $d_{x^2-y^2}$ pairing to a new pairing.

Spontaneous time-reversal-symmetry breaking in the chiral $p$-wave
state can be detected by the muon spin-relaxation measurements
\cite{Sonier} as in Sr$_2$RuO$_4$ \cite{Luke}, or by measuring the
local magnetic field produced by the chiral Andreev surface states.
Quantitative estimates done in Ref.\ \cite{Kwon} show that the latter
effect can be realistically observed with a scanning SQUID microscope
\cite{SQUID}.

\section{Competition between $d$- and $p$-wave pairings}

As discussed after Eq.\ (\ref{bcst}), the ASF interaction $V_s$ enters
in the singlet and triplet superconducting pairing channels with
opposite signs.  Thus, it is unfavorable for $p$-wave pairing, and a
different mediator is needed.  Triplet pairing is usually associated
with the ferromagnetic spin fluctuations, e.g.\ in the superfluid
He--3 \cite{Wolfle} or Sr$_2$RuO$_4$ \cite{Mazin}.  In Ref.\ 
\cite{Kuboki}, the symmetry of superconducting pairing was studied as
a function of the Fermi surface change with doping in a square lattice
model with nearest-neighbor interaction.  It was found that the
symmetry changes with doping from $d$-wave to $p$-wave to
$s$-wave.  The results were applied to Sr$_2$RuO$_4$, but they may be
also relevant to the electron-doped cuprates.

In the calculations given below, we focus on another possible mediator
for $p$-wave pairing, namely the charge-density fluctuations (CDF)
enhanced in the vicinity of the charge-density-wave (CDW) instability.
The role of CDW fluctuations in cuprates was emphasized in Ref.\ 
\cite{DiCastro}.  In a crystal, the CDW wave vector is expected to be
close to ${\bm Q}=(\pi,\pi)$, and the CDF interaction $V_c(q)$ would
have a peak at this vector.  Such interaction has repulsive sign in
the singlet and triplet particle-particle channels, resulting in the
condition (\ref{-Delta}) and supporting both $d$- and $p$-wave
superconducting pairings.

The relative strength of CDF vs.\ ASF in cuprates is subject to
debate, and detailed evaluation of $V_c(q)$ is not the purpose of our
paper \cite{disclaimer}.  Instead, we employ a toy model with the same
interaction in the triplet and singlet channels:
$V_1(q)=V_0(q)=V_c(q)$, where $V_c(q)$ is given by Eq.\ (\ref{V_0}).
Then, the difference in the solutions of the BCS equation (\ref{bcst})
for $d$- and $p$-wave pairings results only from the geometry of the
Fermi surface.  The upper inset in Fig.\ \ref{fig:Delta} shows the
magnitude $|\Delta(\varphi)|$ and the lower inset the phase $\theta$
of $\Delta(\varphi)=|\Delta|e^{i\theta}$ calculated for the chiral
$p$-wave pairing.  We observe that $|\Delta(\varphi)|$ has maxima at
the hot spots angles $\varphi_{\rm hs}$, but, unlike in the
$d_{x^2-y^2}$ case, it does not vanish at $\varphi=\pi/4$ and is not
suppressed when the hot spots approach the zone diagonals.

\begin{figure}[t] 
\includegraphics[width=\linewidth]{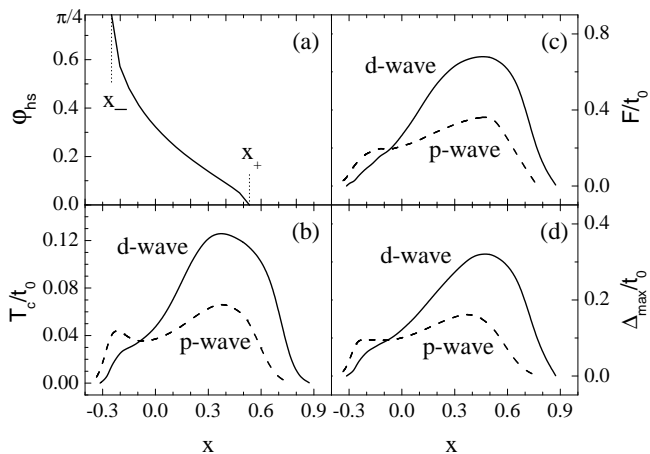}
\caption{Dependence of various quantities on doping $x$.  Panel (a):
  the hot spot angle $\varphi_{\rm hs}$, panel (b): the transition
  temperature $T_c$, panel (c): the condensation energy $F$, and panel
  (d): the maximal gap $\Delta_{\rm max}$.  The solid and dashed lines
  correspond to the $d_{x^2-y^2}$ and to the chiral $p$-wave
  pairings.}
\label{fig:doping}
\end{figure}

In Fig.\ \ref{fig:doping}, we show how various quantities depend on
doping $x$. Panel (a) shows the hot spot angle $\varphi_{\rm hs}$.
Panels (b), (c), and (d) show the transition temperature $T_c$, the
maximal gap $\Delta_{\rm max}$, and the condensation energy $F$ for
the $d_{x^2-y^2}$ and chiral $p$-wave pairings.  It is clear from
Fig.\ \ref{fig:doping} that, at the doping around $x\simeq-8\%$, where
the hot spots approach the zone diagonals closely enough, $p$-wave
pairing wins over $d_{x^2-y^2}$ pairing.  With further increase of
electron doping beyond $x_-$, hot spots disappear, and the proposed
$p$-wave superconductivity rapidly vanishes, in qualitative agreement
with the experimental phase diagram \cite{Uchida,Peng}.  It would be
very interesting to verify this conjecture by ARPES measurements of
the hot spots positions simultaneously with the superconducting phase
diagram in the electron-overdoped regime.

\begin{figure}[t] 
\includegraphics[width=0.9\linewidth]{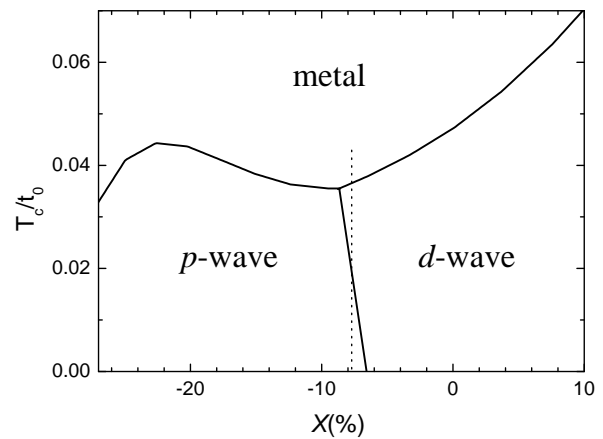}
\caption{Solid lines: Superconducting phase diagram of 
  electron-doped cuprates vs.\ doping $x$ calculated on the basis of
  Fig.\ \ref{fig:doping}.  The vertical dashed line is a guide for
  eye.}
\label{fig:d-p-line}
\end{figure}

Notice that the doping $x_1=-8.8\%$, where the $T_c$ curves for $d$-
and $p$-waves cross in panel (b), is slightly different from the
doping $x_2=-6.6\%$, where the $F$ curves cross in panel (c).  This
means that the critical dopings for the transition from $d$- to
$p$-wave are slightly different at $T_c$ and $T=0$.  Thus, the $d$-$p$
transition line, obtained by connecting the transition points at $T_c$
and $T=0$, is not vertical, as shown in Fig.\ \ref{fig:d-p-line} by
the solid line.  If a sample has the doping $x$ in between $x_1$ and
$x_2$, it should experience a transition from $p$-wave to $d$-wave
with the increase of temperature, as shown in Fig.\ \ref{fig:d-p-line}
by the dashed line.  This effect was actually observed experimentally
in the slightly overdoped samples of $\rm Pr_{1.85}Ce_{0.15}CuO_4$
\cite{Balci-new}.  At low temperature, specific heat was found to
depend linearly on a magnetic field $H$, indicating a fully-gapped
pairing potential consistent with $s$- or $p$-wave.  With the increase
of temperature, the field dependence was found to change to
$\sqrt{H}$, indicating a transition into $d$-wave state, as shown in
Fig.\ \ref{fig:d-p-line}.

In the simplest case, the $d$-$p$ transition line in Fig.\ 
\ref{fig:d-p-line} is the first-order phase-transition line.  Another
possibility, calculated in Ref.\ \cite{Kuboki}, is that this line is
split into two second-order phase-transition lines, and the $p$ and
$d$ phases coexist in the intermediate region.  Which of the two
scenarios takes place is determined by the higher-order coefficients
of the Landau expansion of the free energy (e.g. see discussion in
Ref.\ \cite{Scheven}).  Calculations of these coefficients depend on
fine details of a theoretical model and may be unreliable, e.g.\ they
may be affected by renormalization \cite{Sachdev}.  Thus, the question
of one first-order vs.\ two second-order transitions between $d$- and
$p$-wave phases remains open, both theoretically and experimentally.
We would like to point out that a similar question applies to a
cascade of the magnetic-field-induced phase transitions in observed in
organic conductors \cite{Scheven}.  It was found experimentally
\cite{Scheven} that in high magnetic fields the system exhibits single
first-order phase transitions with hysteresis, whereas in lower fields
it exhibits double-split second-order phase transitions without
hysteresis.  Thus, both scenarios can take place in the same sample
under different conditions.

Positive $x$ in Fig.\ \ref{fig:doping} corresponds to hole doping.  At
$x=x_+$, the hot spots merge and disappear at the van Hove points
$(0,\pi)$ and $(\pi,0)$.  Comparing panels (a) and (b) in Fig.\ 
\ref{fig:doping}, one may notice that the maximum of $T_c$ is achieved
at a hole doping $x<x_+$, and $T_c$ rapidly decreases to zero for
$x>x_+$.  Naively one would expect maximal $T_c$ at $x=x_+$, where the
van Hove singularity is at the Fermi surface.  However, the four hot
spots surrounding each saddle point at $x<x_+$ cover more momentum
space and, thus, produce a higher $T_c$ than at $x=x_+$, where the
four hot spots merge into one.  These results are in qualitative
agreement with the phase diagram of $\rm La_{2-x}Sr_xCuO_4$ mapped to
the ARPES measurements of its Fermi surface in Figs.\ 8 and 7 of Ref.\ 
\cite{Fujimori}.  In the experiment, the maximal $T_c$ is achieved at
$x=15\%$, the Fermi surface passes through the van Hove points at
$x=22\%$, and $T_c$ vanishes at $x=27\%$.  Our theoretical Fig.\ 
\ref{fig:doping} shows the same sequence albeit at different values of
$x$, because our dispersion law parameters $t_0$ and $t_1$ in Eq.\ 
(\ref{sp}) are not optimized for $\rm La_{2-x}Sr_xCuO_4$.

\section{Conclusions}

We have shown that, when the hot spots approach the Brillouin zone
diagonals in the electron-overdoped cuprates, $d_{x^2-y^2}$ pairing
becomes suppressed and is replaced by either singlet $s$-wave or
triplet $p$-wave pairing.  The transition is most likely of the first
order as a function of doping $x$.  To verify the proposed scenario,
it is desirable to measure correlation between superconducting $T_c$
and the hot spots positions by ARPES.  We have given a number of
arguments in favor of the triplet $p$-wave pairing, which may break
the time-reversal symmetry.  The Knight shift measurements in
different samples of electron-doped cuprates show both singlet
\cite{Zheng} and triplet \cite{Zamborszky} superconducting pairing,
which may be an indication of the transition between the two types.
Muon spin-relaxation and the scanning SQUID experiments can detect
spontaneous violation of the time-reversal symmetry.  Relationship
between our proposed theoretical scenario of the superconducting
symmetry change and the phenomenon of the electron dispersion law
flattening is discussed in review \cite{Clark}.

\begin{acknowledgements}
  VMY thanks S. E. Brown and H. Balci for sharing their unpublished
  experimental results \cite{Zamborszky} and \cite{Balci-new}.  VAK
  and VMY thank R.~L.~Greene for useful discussions, and the Kavli
  Institute for Theoretical Physics at Santa Barbara for the
  opportunity to start this collaboration.  VAK and HK thank the
  Condensed Matter Theory Center for arranging their visits to the
  University of Maryland.  VMY is supported by the NSF Grant
  DMR-0137726, HK by the scholarship from Ewha Womans University and
  by Korean Research Foundation, VAK by the NSF Grant PHY-0140316 and
  by the McDonnell Center for Space Sciences, and VAK and MVZ by the
  Grant NS-1885.2003.2 from the Russian Ministry of Industry and
  Science.
\end{acknowledgements}

\emph{Note added in proofs.}  Recently we became aware of Refs.\
\cite{Honerkamp} and \cite{Tremblay}, which studied evolution of
$d$-wave superconductivity in the electron-doped cuprates.



\begin{thebibliography}{00}


\bibitem{Uchida} H. Takagi, S. Uchida, and Y. Tokura, \prl {\bf 62},
  1197 (1989); Nature {\bf 337}, 345 (1989).
  
\bibitem{s-wave} D. H. Wu, J. Mao, S. N. Mao, J. L. Peng, X. X. Xi,
  T.~Venkatesan, R. L. Greene, and S. M. Anlage, \prl {\bf 70}, 85
  (1993).
  
\bibitem{Kirtley} C. C. Tsuei and J. R. Kirtley, \prl {\bf 85}, 182
  (2000).

\bibitem{Kokales} J. D. Kokales, P. Fournier, L. V. Mercaldo,
   V. V. Talanov, R. L. Greene, and S. M. Anlage, \prl {\bf 85}, 3696
   (2000); R. Prozorov, R. W. Giannetta, P. Fournier, and
   R. L. Greene, \prl {\bf 85}, 3700 (2000).
  
\bibitem{Armitage-SC} N. P. Armitage, D. H. Lu, D. L. Feng, C. Kim,
  A. Damascelli, K. M. Shen, F. Ronning, Z.-X. Shen, Y. Onose,
  Y. Taguchi, and Y. Tokura, \prl {\bf 86}, 1126 (2001).

\bibitem{Blumberg} G. Blumberg, A. Koitzsch, A. Gozar, B. S. Dennis,
C. A. Kendziora, P. Fournier, and R. L. Greene, \prl {\bf 88}, 107002
(2002).

\bibitem{Balci} H. Balci, V. N. Smolyaninova, P. Fournier, A. Biswas,
  and R. L. Greene, \prb {\bf 66}, 174510 (2002).
  
\bibitem{Biswas} A. Biswas, P. Fournier, M. M. Qazilbash,
  V. N. Smolyaninova, H. Balci, and R. L. Greene, \prl {\bf 88},
  207004 (2002).
  
\bibitem{Skinta} J.~A.~Skinta, M.-S. Kim, T. R. Lemberger, T. Greibe,
  and M. Naito, \prl {\bf 88}, 207005 (2002).
  
\bibitem{Scalapino} D. J. Scalapino, E. Loh, Jr., and J. E. Hirsch,
  \prb {\bf 34}, 8190 (1986); D.~J.~Scalapino, Phys. Rep. {\bf 250},
  329 (1995).

\bibitem{Pines} P. Monthoux, A. V. Balatsky, and D. Pines, \prl {\bf
    67}, 3448 (1991); \prb {\bf 46}, 14803 (1992).

\bibitem{Ueda} T.~Moriya, Y.~Takahashi, and K.~Ueda,
  J. Phys. Soc. Jpn. {\bf 59}, 2905 (1990).

\bibitem{Yamada} K. Yamada, K. Kurahashi, T. Uefuji, M. Fujita,
  S. Park, S.-H. Lee, and Y. Endoh, \prl {\bf 90}, 137004 (2003).

\bibitem{x-doping} This definition of $x$ is consistent with the
  doping parameter $x$ in the chemical formulas, such as $\rm
  La_{2-x}Sr_xCuO_4$.

\bibitem{Armitage-FS} N. P. Armitage, F. Ronning, D. H. Lu, C. Kim,
  A.~Damascelli, K. M. Shen, D. L. Feng, H. Eisaki, Z.-X.~Shen,
  P.~K.~Mang, N. Kaneko, M. Greven, Y. Onose, Y.~Taguchi, and
  Y. Tokura, \prl {\bf 88}, 257001 (2002).
  
\bibitem{Armitage-PG} N. P. Armitage, D. H. Lu, C. Kim, A. Damascelli,
  K.~M.~Shen, F. Ronning, D. L. Feng, P. Bogdanov, Z.-X.~Shen,
  Y. Onose, Y. Taguchi, Y. Tokura, P.~K.~Mang, N. Kaneko, and
  M. Greven, \prl {\bf 87}, 147003 (2001).

\bibitem{Kusko} C. Kusko, R. S. Markiewicz, M. Lindroos, and
  A. Bansil, \prb {\bf 66}, 140513 (2002).

\bibitem{Dagan} Y. Dagan, M. M. Qazilbash, C. P. Hill, V. N. Kulkarni,
  and R. L. Greene, cond-mat/0310475.
  
\bibitem{disclaimer} The purpose of our calculations is not to achieve
  precise numerical agreement with the experiment, but rather to
  illustrate how the proposed mechanism works within a simple model.
  The parameters of the model are not optimized for any particular
  material, and we neglect their dependence on $x$.

\bibitem{KY} V.~A.~Khodel and V.~M. Yakovenko, Pis'ma ZhETF {\bf 77},
  494 (2003) [JETP Lett. {\bf 77}, 420 (2003)].
 
\bibitem{Guinea} F. Guinea, R. S. Markiewicz, and M. A. H. Vozmediano,
   \prb {\bf 69}, 054509 (2004).

\bibitem{Abrikosov} A. A. Abrikosov, \prb {\bf 53}, R8910 (1996).

\bibitem{Skinta-0.85} J.~A.~Skinta, T. R. Lemberger, T. Greibe, and
  M. Naito, \prl {\bf 88}, 207003 (2002).
  
\bibitem{Peng} J. L. Peng, E. Maiser, T. Venkatesan, R. L. Greene, and
  G. Czjzek, \prb {\bf 55}, R6145 (1997); A. Sawa, M.~Kawasaki,
  H. Takagi, and Y. Tokura, \prb {\bf 66}, 014531 (2002).

\bibitem{Wolfle} D.~Vollhardt and P.~W\"olfle, {\it The Superfluid
    Phases of Helium 3} (Taylor and Francis, London, 1990).
  
\bibitem{symmetry} G. E. Volovik and L. P. Gor'kov, Zh. Eksp. Teor.
  Fiz. {\bf 88}, 1412 (1985) [Sov. Phys. JETP {\bf 61}, 843 (1985)];
  M. Sigrist and K. Ueda, Rev. Mod. Phys. {\bf 63}, 239 (1991).
  
\bibitem{Miyake99} K. Miyake and O. Narikiyo, \prl {\bf 83}, 1423
  (1999).

\bibitem{Sengupta} K. Sengupta, H.-J. Kwon, and V. M. Yakovenko, \prb
  {\bf 65}, 104504 (2002).
  
\bibitem{Zheng} G.-Q. Zheng, T. Sato, Y. Kitaoka, M. Fujita, and
  K.~Yamada, \prl {\bf 90}, 197005 (2003).

\bibitem{Zamborszky} F. Zamborszky, G. Wu, S. Dunsiger, E. Artukovic,
  J.~Shinagawa, A. Lacerda, H. Balci, R. L. Greene, W. G. Clark, and
  S. E. Brown, unpublished.

\bibitem{Ishida98} K. Ishida, H. Mukuda, Y. Kitaoka, K. Asayama,
  Z.~Q.~Mao, Y. Mori, and Y. Maeno, Nature {\bf 396}, 658 (1998).

\bibitem{Chaikin} I. J. Lee, S. E. Brown, W. G. Clark, M. J. Strouse,
  M.~J.~Naughton, W. Kang, and P. M. Chaikin, \prl {\bf 88}, 017004
  (2002).

\bibitem{Sonier} J. E. Sonier, K. F. Poon, G. M. Luke, P. Kyriakou,
  R.~I.~Miller, R. Liang, C. R. Wiebe, P. Fournier, and R. L. Greene,
  \prl {\bf 91}, 147002 (2003).

\bibitem{Luke} G. M. Luke, Y. Fudamoto, K. M. Kojima, M. I. Larkin,
  J.~Merrin, B. Nachumi, Y. J. Uemura, Y. Maeno, Z.~Q.~Mao, Y. Mori,
  H. Nakamura, and M. Sigrist, Nature {\bf 394}, 558 (1998).

\bibitem{Kwon} H.-J. Kwon, V. M. Yakovenko, and K. Sengupta, Synth.
  Met. {\bf 133-134}, 27 (2003).

\bibitem{SQUID} R. C. Black, A. Mathai, F. C. Wellstood, E. Dantsker,
  A.~H.~Miklich, D. T. Nemeth, J. J. Kingston, and J.~Clarke,
  Appl. Phys. Lett. {\bf 62}, 2128 (1993); L.~N.~Vu, M. S. Wistrom,
  and D. J. Van Harlingen, Appl. Phys. Lett. {\bf 63}, 1693 (1993);
  C. C. Tsuei, J. R. Kirtley, C.~C.~Chi, L.~S.~Yu-Jahnes, A. Gupta,
  T. Shaw, J.~Z.~Sun, and M.~B.~Ketchen, Phys. Rev. Lett. {\bf 73}, 593
  (1994); K.~A.~Moler, J. R. Kirtley, D. G. Hinks, T. W. Li, and
  M. Xu, Science {\bf 279}, 1193 (1998).

\bibitem{Mazin} I. I. Mazin and D. J. Singh, \prl {\bf 79}, 733
  (1997); {\bf 82}, 4324 (1999).

\bibitem{Kuboki} K. Kuboki, J. Phys. Soc. Jpn. {\bf 70}, 2698 (2001).

\bibitem{DiCastro} C. Castellani, D. Di Castro, and M. Grilli,
  Z. Phys. B {\bf 103}, 137 (1997).

\bibitem{Fujimori} A. Ino, C. Kim, M. Nakamura, T. Yoshida,
  T. Mizokawa, A. Fujimori, Z.-X. Shen, T. Kakeshita, H. Eisaki, and
  S.~Uchida, \prb {\bf 65}, 094504 (2002).

\bibitem{Balci-new} H. Balci and R. L. Greene, cond-mat/0402263.

\bibitem{Scheven} U. M. Scheven, E. I. Chashechkina, E. Lee, and
  P.~M.~Chaikin, \prb {\bf 52}, 3484 (1995).

\bibitem{Sachdev} M. Vojta, Y. Zhang, and S. Sachdev, \prl {\bf 85},
  4940 (2000).

\bibitem{Clark} J.~W.~Clark, V.~A.~Khodel, M.~V.~Zverev, and
  V.~M.~Yakovenko, Phys. Rep. {\bf 391}, 123 (2004).

\bibitem{Honerkamp} C. Honerkamp, Eur. Phys. J. B {\bf 21}, 81 (2001).

\bibitem{Tremblay} B. Kyung, J.-S. Landry, and A.-M. S. Tremblay,
Phys. Rev. B {\bf 68}, 174502 (2003).


\end{thebibliography}
\end{document}